\documentclass{article}

\usepackage{arxiv}

\usepackage[utf8]{inputenc}  
\usepackage[T1]{fontenc}     
\usepackage{hyperref}       
\usepackage{url}            
\usepackage{booktabs}       
\usepackage{amsfonts}       
\usepackage{nicefrac}       
\usepackage{microtype}      
\usepackage{lipsum}
\usepackage{cite}
\usepackage{graphics}
\usepackage{graphicx}

\title{Eliminating Systematic Bias from Difference-in-Differences Design: A Permutational Detrending Strategy}

\author{
  Xiaoming Wang\thanks{Corresponding author. Health Services Statistical and Analytic Methods, Alberta Health Services, \#7235, 2nd Floor, West Wing, Aberhart Centre, 11402 University Avenue, Edmonton, AB T6G 2J3.} \\
  Health Services Statistical and Analytic Methods\\
  Alberta Health Services\\
  Edmonton, AB T6G 2J3 \\
  \texttt{xiaoming.wang@ahs.ca} \\
   \And
 Sukun Wang \\
  Financial Crime Unit\\
  Bank of Montreal\\
  Toronto, ON, M4W 1A6, Canada \\
  \texttt{sukun492@ualberta.ca} \\
}

\begin{document}
\maketitle

\begin{abstract}
Since the initial work by Ashenfelter and Card in 1985, the use of difference-in-differences (DID) study design has become widespread. However, as pointed out in the literature, this popular quasi-experimental design also suffers estimation bias and inference bias, which could be very serious in some circumstances. In this study, we start by investigating potential sources of systemic bias from the DID design.  Via analyzing their impact on statistical estimation and inference, we propose a remedy -- a permutational detrending (PD) strategy -- to overcome the challenges in both the estimation bias and the inference bias. We prove that the proposed PD DID method provides unbiased point estimates, confidence interval estimates, and significance tests. We illustrate its statistical proprieties using simulation experiments. We demonstrate its practical utility by applying it to the clinical data EASE (Elder-Friendly Approaches to the Surgical Environment) and the social-economical data CPS (Current Population Survey). 
\end{abstract}

\section{Introduction}
Observational studies are commonly adopted to evaluate the impact of policy changes. One limitation in using a pre-and-post design is the need for control for the underlying time trend\cite{bib1}. To overcome the bias issue from an underlying time-trend, Ashenfelter and Card in 1985 proposed the difference-in-differences (DID) study design\cite{bib2}. 

Since the initial work by Ashenfelter and Card, the use of the DID study design has become widespread\cite{bib3,bib4, bib5, bib6, bib7, bib8}. For empirical research, the DID design was supposed to be unbiased and powerful to distinguish between effective and ineffective policies. However, as pointed out in the literature, the DID study design also suffers a bias issue of over-rejection in estimation, which could be very serious in some circumstances\cite{bib7, bib8}. 

Bertrand et al.\cite{bib8} were among the surprising few who scrutinized the discriminatory ability of the DID design. Imposing effective and ineffective interventions (or `laws' in their paper) by simulation on weekly earning data from the Current Population Survey (CPS), they found that estimation from a DID analysis was biased in size of a significance test. They then proposed methods using data collapsing, block bootstrapping, and clustered standard error estimation.  They believed that it was the correlation among observations that leads to the over-rejection rates. Followed their work, researchers have evolved different strategies to reduce estimation bias in standard error (SE). Although lots of efforts have been made and some progress has been achieved over the last two decades. However, the performance of those strategies is still disappointing\cite{bib7}. We will provide more details in subsection 2.3. 

On the other hand, the bias issue in the point estimate of an intervention effect in a DID analysis has received scant attention in the literature. In practice, it is implausible to select reference groups to be ``perfectly" matching the intervention groups in their underlying time trends. If the parallel trends assumption fails, the point estimate from a DID analysis will be biased. Also, the uncontrolled part of time trends will enter into the model errors, leading to a potential higher level of error correlation. We believe that the violation of the parallel trends assumption is the dominant source of systemic bias. We argue that focusing on the inference bias and ignoring the estimation bias could be misleading. In literature, we haven't seen any promising solution to both the bias in point estimate and the bias in SE estimate. We must seek an appropriate solution to the systematic bias challenge by scrutinizing overall bias sources and their impacts.

In this paper, we investigate potential sources of systematic bias in the DID design. We propose an appropriate solution to the challenges in both the estimation bias and the inference bias. It provides unbiased estimates and unbiased references for intervention effects under linear models. Also, it can be naturally generalized to other potential models used for DID analyses and keeps the merits.  

The remainder of this paper proceeds as follows. Section 2 explores potential sources of systematic bias from the DID design and analyzes their impact on statistical estimation and inference. Section 3 proposes a detrending strategy to handle the estimation bias and then a permutational detrending (PD) strategy to eliminate the inference bias. We establish asymptotic unbiasedness of the proposed approaches. Section 4 illustrates their statistical proprieties using simulation experiments. We also demonstrate their practical utility by applying it to the EASE data and the CPS data. We discuss the strengths and limitations of the proposed approaches in section 5.

\section{Potential Sources of Bias}

\subsection{A brief on the DID design}
In a DID study, we measure an intervention effect by looking at the difference between the pre-and-post effects in the intervention and reference groups\cite{bib2}. Formally, let $Y_{igt}$ be the outcome of interest for individual $i$ in group $g$ at time $t$.  We try to see the effect of an intervention $I_{gt}$ (a dummy variable indicating whether the intervention affected group $g$ at time $t$). Let's consider, for simplicity of illustration, a linear regression model
\begin{equation}
Y_{igt}=\alpha_0+\alpha A_g+\beta B_t+\gamma I_{gt}+ \varepsilon_{igt}    
\end{equation}
where the three main regressors represent fixed effects of group $g$ ($A_g=1/0$ indicating whether the intervention affected group $g$), time $t$ ($B_t=1/0$ indicating whether the intervention implemented at time $t$), and their interaction ($I_{gt}=A_g\times B_t$), respectively. $\gamma$ is the intervention effect interested. Status of group $g$ and time $t$ could be multiple. For each individual $i$, there is an unique group statue $g$ (either $g$ in intervention groups $\mathcal{I}$ or reference groups $\mathcal{R}$) but possible multiple time points $t^i_1, \cdots, t^i_{n_i}$ in single time period or multiple time periods. 

Denoting by $\alpha_1=\alpha_0+\alpha$ and $\beta_1=\beta+\gamma$, model (1) for intervention groups can be written as \begin{equation}
Y_{igt}=\alpha_1+\beta_1 B_t+\varepsilon_{igt}, \quad g\in\mathcal{I}.    
\end{equation}
Denoting by $\beta_0=\beta$, model (1) for reference groups can be written as \begin{equation}
Y_{igt}=\alpha_0+\beta_0 B_t+\varepsilon_{igt}, \quad g\in\mathcal{R}.    
\end{equation}
The intervention effect $\gamma$ is the difference between $\beta_1$ and $\beta_0$. 

\subsection{Potential sources of systemic bias} 
A basic assumption of the DID approach is the parallel trends assumption, which supposes that {\it in the absence of intervention effect, outcomes of the intervention and reference groups would follow parallel paths over time}\cite{bib1}. Under this assumption, any temporal factors other than the policy is under control. It allows the DID design to account for unobserved temporal variables. A key to implementing the DID design is to find reference groups for which the parallel trends assumption holds. Ideally, the only difference between the intervention and the reference groups would be the exposure to the policy. However, such ``perfect" reference groups may be difficult or even impossible to find, and violation of the parallel trends assumption could lead to estimation bias of the intervention effect. 

In practice, temporal factors including long-term time-trends, seasonal time-trends, and short-term shocks may confound the estimate of intervention, for their potential correlation with the exposure $B_t$ (which is also a time-varying variable). Time-invariant factors, such as individual characteristics (gender, race, social-economic status, etc.), could also raise estimation bias. It is because observations are not from the same group of individuals with time going. These factors are constant for each individual.  But they are time-varying with changing cohort of individuals under observation, and hence could be correlated with $B_t$ as well. If the parallel trends assumption fails, a lack of sufficient adjustment could result in bias in the point estimate of the effect. 

Potential systematic bias could also come from the used estimation equation (or likelihood function), leading to a biased SE estimate and a biased significance test. This situation is similar to that of overlap bias in a matched case-control study\cite{bib9} or a case-crossover study designs\cite{bib10, bib11}. We have generally recognized that, with structured data, a modeling method designed only for independent observations will provide a biased estimate of SE and a biased significance test. Clustering within groups and repeated measures of individuals could lead to correlation among observations.  Uncontrolled part of the time trends too if the parallel trends assumption fails.

\subsection{Limitation of bias reduction methods in the literature}
Much effort has been made in the literature to overcome the challenges of bias in a DID study. The main focus was on reducing bias in SE estimates from within-group correlation in outcomes. It was for more accurate size and possible higher power in significance tests. As reviewed by Rokicki et al.\cite{bib7}, the approaches used to account for within-group correlation in outcomes can be divided into three broad categories: (1) post hoc adjustments such as CSE (Clustered Standard Errors)\cite{bib12}, bootstrapping\cite{bib13, bib14, bib15}, or permutation tests\cite{bib16, bib17, bib18, bib19, bib20}; (2) explicitly modeling the within-cluster error correlation\cite{bib21, bib22, bib23, bib24} such as GEE, random-effect models, and feasible generalized least squares; and (3) aggregating the data to the group level, thereby eliminating the within-group correlation\cite{bib8, bib25}. For more details of these approaches, we refer readers to the review paper of Rokicki et al.\cite{bib7} and related references therein.

The idea of aggregating data to group level was proposed by Bertrand et al.\cite{bib8} for repeated cross-sectional data. It is most commonly used for economic outcomes such as income or hours worked. However, this idea may not be suitable to be generalized to unbalanced data for evaluating a group-level policy on individual-level outcomes. Also, it works fare poorly as the number of groups gets small\cite{bib7}. Additionally, data aggregation unavoidably leads to information loss and results in lower efficiency of inference.

In practice, it is difficult to look into the covariance structure of errors and correctly specify it in a model (like GEE or a random-effect model). We need much more information to estimate the covariance matrix because of its high dimension. Additionally, considering a possible estimation bias in point estimate from confounding, all the strategies focus on explicitly modeling the within-cluster error correlation could be less efficient. Their performance could be getting even worse as the sample size goes down.

Some of the post hoc adjustments, such as the approximate permutation procedure\cite{bib20}, could be promising for handle within-group correlation. However, it is sensitive to the parallel trends assumption. The reason is simple and direct: even though one can reveal the null distribution of an intervention effect, the estimation bias in the point estimate also leads to biased reference. We will see in section 4 that `placebo' intervention also leads to over-rejection rates as the parallel trends assumption fails. 

Although some progress has been made on control the bias in SE estimation, little attention has been paid to the bias in point estimates. It is the main limitation of the approaches motioned above in this subsection.

When the parallel-trend assumption fails, some authors resort to a polynomial trend-augmented version of the original DID model in their application studies\cite{bib26}. Vandenberghe \cite{bib26} proposed a new method using pre-intervention observations to capture linear or non-linear trend differences. Using a Monte Carlo simulation experiment, Ryan et al. \cite{bib27} tested the performance of several estimators when the parallel trends assumption fails. These estimators were original DID, DID with propensity score matching, single-group interrupted time-series analysis, and multi-group interrupted time-series analysis. Leavitt \cite{bib28} innovated new methods to handle both the estimation bias and inference bias simultaneously. The main idea of the methods was to predict the counterfactual outcomes in the absence of intervention. The author provided two unbiased estimators (under conditions) and developed an empirical Bayesian procedure for the inference. However, the proposed methods were limited to the canonical DID linear model without adjustment for covariates. How to generalize them to a general model setting is still under question.

To our knowledge, none of the methods in the literature (1) copes with both the estimation bias and the inference bias simultaneously, and (2) can be generalized for all the potential models used for a DID analysis. Applications need an ideal solution to eliminate systemic bias for all potential models adopted for a DID analysis, including the generalized linear models. The method we show in the next section is just the one expected.

\section{A Permutational Detrending Strategy}
\subsection{Detrending DID analysis}
Without the parallel trends assumption, we have to consider adjustments for various covariates. Under this circumstance, we can rewrite model (1) as
\begin{equation}
Y_{igt}=\alpha_0+\alpha A_g+\beta B_t+\gamma I_{gt}+ \lambda_g X_{igt}+\varepsilon_{igt},    
\end{equation}
where $\lambda_g X_{igt}$ represents effect from all other potentially covariates, including individual characteristics and temporal impact factors other than the intervention. Using model (4) in practice could be very difficult because of the complexity of a large amount of potentially covariates, both observed and unobserved. 

By separating the time-trend effect from the others, the following alternative model could be more practically useful:
\begin{equation}
Y_{igt}=\alpha_0+\alpha A_g+\beta B_t+\gamma I_{gt} + \lambda_g (t)+\mu Z_{i}+ \varepsilon_{igt},    
\end{equation}
where $\lambda_g(t)$ represents the underlying time trend of group $g$ -- reflecting impact from temporal factors, and $\mu Z_{i}$ represents the effect from the characteristics of individual $i$. It is obviously to see that $Y_{igt}-\lambda_g (t)$ can be viewed as a detrending response generated from a traditional DID model. 

We suggest using a simple version of the model (5)
\begin{equation}
Y_{igt}=\alpha_0+\alpha A_g+\beta B_t+\gamma I_{gt}+ \lambda_g t+\mu Z_{i}+\varepsilon_{igt}.
\end{equation}
Where $\lambda_g t$ represents the linear part of the underlying time-trend of the group $g$. In practice, underline time trends could be non-linear. Here we consider only their linear components for simplicity.\footnote{One can also consider fully detrending (both the linear and non-linear time trends) by using polynomial or cubic spline technique.} In this case, their non-linear parts will enter the model errors.

Model (6) is much simpler than model (4) for both theoretical and practical studies. It suggests considering a linearly detrending of $Y_{igt}$ (either by subtracting underline time-trends $\lambda_g t$ from $Y_{igt}$, or equivalently by model adjustment for them). We call it the detrending DID analysis method. We will see in our simulation study that the detrending technique functions as the core to eliminate systematic bias in a DID study. 

If the parallel trends assumption fails, we can overcome the challenge of bias in point estimates by directly fitting data using the detrending DID model (6). Suppose model (6) holds. Under some regular conditions, for example, 
\begin{equation}
E\varepsilon_{igt}=0 \mbox{ and } \varepsilon_{igt} \mbox{ is independent with the regressors} , 
\end{equation}
the detrending DID model (6) provides unbiased point estimate for the intervention effect $\gamma$.\footnote{Estimate from logistic regression or other generalized models, can be asymptotic unbiased under some regular conditions.} 

\subsection{Permutational detrending DID analysis}
We show another technique to eliminate inference bias in a DID analysis. Our strategy is simply doing permutation on the individuals' records $(Y_{igt}, Z_i)$ in the model (6). Because this technique combines both detrending and permutation, we call it permutational detrending DID analysis or acronym PD DID. We have the following algorithm for the proposed method. 

\noindent\textbf{Algorithm: permutational detrending DID analysis}
\begin{enumerate}
\item[1.] Estimate the intervention effect $\hat{\gamma}$ (as point estimate) using model (6) and the original data for a DID study;

\item[2.] Permute randomly $\{(Y_{igt},Z_i),g\in\mathcal{I}\}$ and $\{(Y_{igt}, Z_i),g\in\mathcal{R}\}$ separately,\footnote{After permuting, matching between an observation and an observing time will be random.} and estimate the intervention effect $\hat{\gamma}^*$ using model (6) and the data after permuting;

\item[3.] Create empirical distribution of estimation bias $\mathcal{D}=\{\hat{\gamma}^*_j|\quad j=1,\ldots, M\}$ by independently replicating above experiment (step 2) $M$ times;

\item[4.] Calculate the mean $\bar{\gamma}^*=\sum_{j=1}^M \hat{\gamma}^*_i/M$, and the 95\% confidence interval $(\hat{L}^*,\hat{U}^*)$ of the empirical distribution;

\item[5.] Adjust $(\hat{L}^*,\hat{U}^*)$ to $(\hat{L},\hat{U})$ =$(\hat{L}^*+\hat{\gamma},\hat{U}^*+\hat{\gamma})$ (as 95\% confidence interval estimate);

\item[6.] The significance of the intervention effect (p-value) can be measured by the rank of $\hat{\gamma}$ in the empirical distribution $\mathcal{D}$.
\end{enumerate}

To build an empirical null distribution of $\hat{\gamma}$ is a crucial step to the PD DID approach. We suggest using a relatively large number of replicates (e.g., $M \ge 500$) in step 2. In the following simulation and application studies, we set $M=1,000$ to get a relatively reliable distribution and relatively high resolution of $p$ value. 

If the model errors in (6) are correlated, then the traditional SE estimate for $\gamma$ is biased. In this case, the traditional statistical inference is inappropriate. Alternatively, we choose to compare $\hat{\gamma}$ with those $\hat{\gamma}^*$s in $\mathcal{D}$. We will prove that the PD DID analysis provides an unbiased significance test.

\subsection{Asymptotic property of the PD DID Strategy}
Theoretically speaking, if the model (6) and the condition (7) hold, then the PD DID analysis provides unbiased point estimates, confidence interval estimates, and significance tests for $\gamma$, as permutation replication times $M\rightarrow\infty$. Our reasoning is as follows. 

Under condition (7), model (6) provides unbiased estimates for all the parameters. So is the point estimate for  $\gamma$ from the PD DID analysis. Because the confidence interval and the significance test for $\gamma$ depend on the estimated null distribution $\mathcal{D}$, we will focus on proving that the empirical distribution $\mathcal{D}$ is asymptotically unbiased as permutation replication times $M\rightarrow \infty$.

Let's start with a stronger null hypothesis: individual's record $(Y_{igt}, Z_i)$ is independent with time $t$. It implies that, $\gamma=0$, $\beta=0$ and $\lambda_g = 0$ in model (6). Under the stronger null hypothesis, all the $M$ permuted samples $\{(Y^{*}_{igt}, Z^{*}_i)\}_j$, $j=1,\ldots, M$, are independent and identically distributed replications of the original sample $\{(Y_{igt}, Z_i)\}$. It implies that the parameter estimates under permuted samples $\{(\hat{\gamma}_j^*,\hat{\beta}_j^*,\hat{\lambda}_{gj}^*)$, $j=1,\ldots, M\}$ are also independent and identically distributed replications of the parameter estimates under the original sample $(\hat{\gamma}_j, \hat{\beta}_j,\hat{\lambda}_{gj})$. Therefore, the empirical distribution $\mathcal{D}$ approaches the true distribution of $\hat{\gamma}$ under the null hypothesis, as $M\rightarrow\infty$. It guarantees asymptotic unbiasedness of the confidence interval estimate and significance test.\footnote{This conclusion is not limited to a linear regression model. It holds for all models adopted for a DID analysis if and only if the point estimate is unbiased. If the point estimate is asymptotically unbiased as the sample size $N\rightarrow\infty$, then the confident interval estimates and the significance test is asymptotically unbiased as both $N$ and $M\rightarrow\infty$.} 

\section{Simulation and Empirical Studies}

\subsection{Simulation design}
The simulation experiments in this paper are designed for (a) bias check and size check under the null hypothesis and (b) bias check and power check under alternative hypotheses, for the original DID analysis, the detrending DID analysis, and the permutational detrending DID analysis.

Simulation data are generated from the following data generating process.
\begin{equation}
Y_{igt}=\gamma I_{gt}+\lambda_g t+\alpha_g+u_g+v_i+w_{it}    
\end{equation}
with 
$$u_g\sim N(0,\sigma_u^2); \quad v_i\sim N(0,\sigma_v^2);$$
$$w_{it}\sim \mbox{AR(1) with } N(0, \sigma_w^2) \mbox{ distribution}.$$
Where the first three items represent fixed effects. $\gamma$ is the true effect of the intervention. $\lambda_g t$ represents the underlying time-trend in the group $g$. And $\alpha_g$ is the mean effect of the group $g$. The last three items are random effects. $u_g$ represents random effect at group-level. $v_i$ represents the random effect at the individual level, and $w_{it}$ represents the random effect at the observation level. Pare-wise correlation among within-group individual effects is set at $\rho$ (i.e., $COR(v_i,v_j)=\rho$ for individuals $i$ and $j$ in group $g$).  Observations from the same individual are viewed as repeated measures and generated by a first-order auto-regression (AR(1)) process with normal distributions and one autocorrelation parameter $\rho$. The AR(1) process allows observations from each individual to be serially correlated (i.e., $COR(w_{it}, w_{i,t-1})=\rho$). Via this data generating progress, the observations within-group and within-individual will be correlated if $\rho > 0$, and are independent if $\rho = 0$.

In the following simulation experiments, we consider two intervention groups and two reference groups. Each of the four groups has $n=200$ individuals, and each individual has $1$ to $7$ observations (with $4$ observations on average for each individual). We set the duration of the study as one year (or $365$ days), the first half-year (182 days) as the pre-intervention period, and the second half-year (183 days) as the post-intervention period. We randomly select a visit/observing date for each of these observations ($Y_{igt}$, totally 3,200) among the 365 days. 

For simulation 1 and 2 in the next subsections, we set $\alpha_g= \pm 0.5$ for intervention/reference groups and $\sigma_u=0.1$ (for the group-level effect $u_g$), $\sigma_v=1$ (for the individual-level effect $v_i$),  $\sigma_w=0.1$ (for the AR(1) process). We let $\gamma$, $\lambda_g$ and $\rho$ change to create different scenarios.\footnote{The setting of the parameters (the first setting) is quite empirical. To see influence of a setting on the findings, we also try another setting (the second setting): set ($\alpha_g=\pm 5$, $\sigma_u=1$, $\sigma_v=1$, and $\sigma_w=1$) and let $\gamma$, $\lambda_g$ and $\rho$ change to create different scenarios. Corresponding results are reported in the supplementary Excel file ``S-Table 1". We get the same conclusions from the results, excepting relatively lower power of significance test (because the signal-noise ratio is relatively lower under the second setting). It implies that the findings reported in the paper is not associated with the setting of parameters.}

\subsection{Bias check and size check under null}
We try by the simulation to check the bias in point estimate and the size of significance test for the three candidate analysis methods. \\

\noindent \textbf{Simulation 1.} Set intervention effect $\gamma=0$, and $\rho=0$, $0.5$ or $0.9$, and underlying time-trend slop $\lambda_g=\pm l/365$ for intervention/reference groups, with the time-trend slop parameter  $l$ takes a value from the following list $\{-0.5, \ldots, -0.1, 0, 0.1, \ldots, 0.5\}$. We replicate 1,000 experiments for each setting of $(\gamma, \rho, l)$. In each experiment, we simulate data using the data generating process. Then the original DID analysis method and the proposed detrending DID analysis approaches are adopted to fit the data. The average value of the 1,000 point estimates and the reject frequency (with significance criteria 0.05) in the 1,000 significance tests for intervention effect $\gamma$ are reported in Table 1. 

\begin{center}
   \mbox{(Insert Table 1 near here)}
\end{center}

From the table, we see that, under the null hypothesis ($\gamma=0$) and also under the parallel trends assumption ($l$=0), the bias in point estimate from the original DID analysis is very close to zero, and the size of a significance test is close to the nominal size 50 in 1,000 replications. Unfortunately, the bias of the estimated effect also goes up (down) as the time-trend slop parameter $l$ goes up (down), leading to higher and higher rejection rates. It implies that the original DID analysis heavily depends upon the parallel trends assumptions. In other words, violation of the parallel trends assumption could lead to unacceptable bias in both point estimate and significance test. We detected no significant difference between the results from $\rho=0$, $\rho=0.5$ and $\rho=0.9$ scenarios. It suggests that the dominating source of bias in these simulated scenarios roots from the violation of parallel trends assumption. 

Just as we expected, the detrending DID analyses eliminate both biases in the point estimate and the significance test in all the simulated scenarios under the null hypothesis. Comparing the performance of the original DID analysis with that of the detrending DID analysis (or the PD DID analysis), we conclude that it is an insufficient adjustment for time-trend that leads to a biased point estimate and a biased significance test. Under these simulated scenarios, we can't see any significant improvement from the PD DID analysis over the detrending DID analysis. 
\subsection{Bias check and power check under alternatives}
Creating scenarios under alternative hypotheses, we do a bias check and a power check for the three candidate analysis methods.

\noindent \textbf{Simulation 2.} This simulation is similar to Simulation 1, except setting $\rho=0.5$, the intervention effect $\gamma$ taking a value from $\{0, 0.05, \ldots, 0.5\}$ and the time-trend slops $\lambda_g=\pm l/365$ for intervention/reference groups with the slop parameter  $l=-0.2/0/0.2$. At each setting of $(\gamma, \rho, l)$, the average value of the 1,000 point estimates and reject frequency (with significance criteria 0.05) in the 1,000 significance tests for the intervention effect $\gamma$ are depicted in Figure 1.

\begin{center}
   \mbox{(Insert Figure 1 near here)}
\end{center}

The left column of Figure 1 explains the performance of the original DID analysis. It tells us again the biased nature of the original DID design as the parallel trends assumption fails. From the upper-left panel of Figure 1, we see that the influence on point estimate from time-trend is nearly constant. In detail, if the time-trend parameter $l$ is $0/0.2/-0.2$, then estimation bias is also close to $0/0.2/-0.2$. We can find the same phenomenon in Simulation 1 (see Table 1). The bottom-left panel shows that the power of the original DID design is perfect when the parallel trends assumption holds and misleading as it fails.

Contracting to the original DID analysis, our detrending DID analysis approach tells a different story (the middle column of Figure 1). Whenever the parallel trends assumption holds or not, the proposed detrending DID analysis provides an accurate point estimate of the intervention effect (the upper-middle panel of Figure 1). It implies that the proposed detrending approach provides an unbiased point estimate under alternatives. Also, the proposed detrending DID analysis provides an unbiased significance test for the intervention effect (bottom-middle panel of Figure 1). It is ``independent" with the underlying time-trends in both the intervention and reference groups. Again, the PD DID analysis performs similarly to the detrending DID analysis (right column of Figure 1).

Under the parallel trends assumption, the original DID analysis offers the best test power. However, it is sensitive to the underlying time trends. While the two detrending DID approaches provide unbiased and robust results (no matter the parallel trends assumption holds or not), but with relatively lower test power. Gaining unbiasedness using the detrending strategy is not free. We need data information to estimate the time trends in the intervention and reference groups. 

\subsection{An empirical study on EASE data}
Using clinical data, we illustrate applying the proposed methods to avoid biased estimates in DID design studies. The data were from the Elder-Friendly Approaches to the Surgical Environment (EASE)  study \cite{bib29}. Briefly, It was a prospective, non-randomized, controlled pre-and-post study at two tertiary care hospitals (University of Alberta Hospital, Edmonton, and Foothills Medical Centre, Calgary) in Alberta, Canada, from April 14, 2014, to March 28, 2017. The EASE program was a surgical quality improvement initiative designed for older patients (especially those with frailty) in an emergency surgical setting. 

Data were collected before the intervention from April 14, 2014, to July 23, 2015. It was followed by a 3-month implementation period after introduced the EASE initiative to the Edmonton site. The data used in the empirical study contains 684 records from the Edmonton side (153 and 140 in pre and post periods) and Calgary side (169 and 222 in pre and post periods). We extract two dummy variables (indicating intervention side and post-intervention period), surgery start date, four outcome variables (in-hospital death, postoperative serious infections in hospital, primary outcome in hospital, and total length of stay (LOS) in hospital), and baseline characteristics of the patients (age, sex, ASA status, surgery type, and Charlson comorbidity index).\footnote{The primary outcome was a composite of major postoperative in-hospital complications or death. The ASA status represents the American Society of Anesthesiologists [ASA] physical status classification.} For more details, we refer readers to the EASE study paper\cite{bib29}.

\begin{center}
   \mbox{(Insert Table 2 near here)}
\end{center}

The three binary outcomes were analyzed using logistic regression. The duration outcome (total LOS) was analyzed using negative binomial regression. All the models were adjusted with the five characteristics of the patients. We report point estimates of the intervention effect and corresponding p-values of significance test in Table 2. From the table, we see the following.  First, the detrending  DID makes a significant change in the estimated effects from the original DID, which indicates non-ignorable biases in estimates from the original DID analyses.  Second, the permutational detrending DID provides similar p-values comparing to the detrending DID on the three binary outcomes but not on total LOS, which implies the necessity of handling bias in SE estimate in practice sometimes. Third, results from the detrending DID analyses indicate that the EASE intervention reduced the in-hospital mortality risk but was far from significant. 

\subsection{An Empirical study on CPS data}

The CPS data is one of the most popular used data in DID literature. We provide this empirical study to answer an important question: where does systematic bias root in social-economical police analyses using DID designs? The focus is on illustrating methodology utility instead of exploring a social-economical policy issue. 

We extract data on women in their fourth interview month in the Merged Outgoing Rotation Group of the CPS from 1979 to 1999. There are 549,735 observations of women between 25 and 50 years old who lived in the United States, which provide information on weekly earnings, employment status, education, age, and state of residence. We define wage as log(weekly earnings) as the dependent variable and divide the study period into pre (1979-1989) and post (1990-1999) periods. We select from the 50 states the top 12 in size of observations (California, New York, Texas, Ohio, Illinois, Florida, Pennsylvania, Michigan, New Jersey, North Carolina, Massachusetts, and Virginia). For each pair among the 12 states (total 66 combinations), we select the first one as `intervention' and the second as `reference.' We estimate the interactive effect between pre-and-post and intervention-and-reference, using the original DID, detrending DID, and permutational detrending DID analyses. We include age and education as covariates in each of the 198 DID models. Supplementary material (S-Table 2) contains the results of the three DID analyses. Estimates and p-values by analysis methods are depicted in Figure 2. 

\begin{center}
   \mbox{(Insert Figure 2 near here)}
\end{center}

Estimates before-and-after detrending are different (the top panels of Figure 2). The correlation between the two groups of point estimates is 0.504. It implies that the parallel trends assumption fails for the outcomes in most of the 66 pairs of states. The difference in point estimates also leads to a big difference in reference conclusions (the bottom panels of Figure 2). Among the 66 pairs of states, 31 (47.0\%) and 17 (25.8\%) intervention effects are reported as significant (p-value $<0.05$) before-and-after detrending, respectively. The correlation between the two groups of p-values is close to zero (-0.046). To our surprise, the p-values before-and-after permutation are very similar (the bottom panels of Figure 2). The correlation between the two groups of p-values is close to one (0.997). It indicates the dominant systematic bias is from unparalleled time trends in most of these original DID studies.  On the other hand, the impact from correlation among observations is trivial.

\begin{center}
   \mbox{(Insert Table 3 near here)}
\end{center}

Table 3 contains the modeling results with significant effect after Bonferroni adjustment (p-value $<0.05/66$) estimated by the detrending DID or permutational detrending analyses. Under this criterion, we find three pairs of states with significant interaction effects between pre-and-post and intervention-and-reference. We believe there was a big difference between pre-and-post policy changes in the three pairs of states.

\section{Discussion}
In a DID analysis, a difference between underlying time trends results in a bias in point estimate, while error structure leads to a bias in SE estimate. Under the parallel trends assumption, the original DID analysis provides an unbiased point estimate and a powerful significance test. It becomes complicated, however,  as this basic assumption fails.

Detrending provides an unbiased point estimate for the intervention effect. It is the key to eliminate systematic bias in a DID study. Without it, we can't do permutation. It is because the matching between individuals' observations and corresponding observing times is unexchangable. With detrending, the permutation algorithm can control inference bias from correlations among errors. We've proved in linear regression that: (1) the detrending DID analysis provides an unbiased point estimate for the intervention effect, and (2) the permutational detrending DID analysis provides both an unbiased point estimate as well as an unbiased significance test. It is worth noting that the contribution of the permutation algorithm is to create an unbiased estimate of the null distribution (as permutation replication times $M\rightarrow\infty$). We see from the proving process that it is model-free and distribution-free. 

Our simulation experiments support the theoretical conclusions. We see from them that: (1) after detrending, both biases in point estimate and the significance test are gone; and (2) the function of the permutation algorithm is trivial on bias in the significance test. We guess that the over-rejection issue in DID analyses in the literature was mainly from bias in point estimate instead of bias in SE estimate. Our empirical studies in both clinical and social-economical research areas support the conclusion. Because of this, we believe that detrending is the top crucial step to a DID analysis. We argue that considering bias in SE and ignoring bias from underlying time trends could be misleading. 

The advantage of the detrending technique is at achieving unbiasedness of a point estimate but with a loss of efficiency in a significance test. The advantage of the permutation algorithm is at getting an unbiased estimate of the null distribution but with a computational burden. In practice, we can use one or both based on the need of an application. If the parallel trends assumption holds, we don't need to use detrending. We can employ the permutation algorithm to handle a potential correlation between errors without losing any efficiency on the significance test. The detrending technique and the permutational detrending algorithm are simple and effective. It is easy to apply them to any model selected for a DID analysis, such as a generalized linear model.

\section*{Acknowledgements}
We thank professor Khadaroo R.G. for her permission to use the EASE data. This study was approved by the Health Research Ethics Board at the University of Alberta and the need for patient-level informed consent was
waived. No patient records/information was required or identified. 

\subsection*{Funding}
This research received no specific grant from any funding agency in the public, commercial, or not-for-profit sectors.

\subsection*{Conflicts of Interest}
The author(s) declared no potential conflicts of interest with respect to the research, authorship, and/or publication of this article.

\subsection*{Supplementary Material}
{\it Supplementary material 1}: S-Table 1. Simulation results under the second setting of parameters.\\
{\it Supplementary material 2}: S-Table 2. Results of application study on CPS data.\\

\newpage
\begin{table}
 \caption{Size check and bias check for original DID, detrending DID, and permutational detrending DID analyses.}
  \centering
\begin{tabular}{cccc|lll|lll|ll}
\toprule
\multicolumn{3}{c}{Scenario}&&\multicolumn{2}{c}{Original DID}&&\multicolumn{2}{c}{Detrending DID}&& \multicolumn{2}{c}{PD DID}\\
\midrule
$\gamma$ & $l$ & $\rho$ && Size & Bias  && Size & Bias  && Size & Bias \\
\midrule
0 & -0.5 & 0.0 && 1000 & -0.502 && 49 & -0.007 && 46 & -0.007 \\
0 & -0.4 & 0.0 && 1000 & -0.399 && 56 & 0.006 && 54 & 0.006 \\
0 & -0.3 & 0.0 && 982 & -0.299 && 42 & 0.005 && 41 & 0.005 \\
0 & -0.2 & 0.0 && 808 & -0.199 && 56 & -0.003 && 59 & -0.003 \\
0 & -0.1 & 0.0 && 283 & -0.099 && 64 & 0.005 && 60 & 0.005 \\
0 & 0.0 & 0.0 && 68 & -0.003 && 55 & -0.008 && 54 & -0.008 \\
0 & 0.1 & 0.0 && 263 & 0.094 && 53 & -0.001 && 60 & -0.001 \\
0 & 0.2 & 0.0 && 812 & 0.200 && 37 & -0.001 && 41 & -0.001 \\
0 & 0.3 & 0.0 && 985 & 0.298 && 51 & -0.004 && 49 & -0.004 \\
0 & 0.4 & 0.0 && 1000 & 0.402 && 47 & -0.002 && 40 & -0.002 \\
0 & 0.5 & 0.0 && 1000 & 0.501 && 51 & 0.002 && 50 & 0.002 \\
\hline
0 & -0.5 & 0.5 && 1000 & -0.501 && 59 & 0.002 && 55 & 0.002 \\
0 & -0.4 & 0.5 && 1000 & -0.398 && 55 & 0.003 && 56 & 0.004 \\
0 & -0.3 & 0.5 && 993 & -0.300 && 41 & 0.006 && 41 & 0.006 \\
0 & -0.2 & 0.5 && 881 & -0.196 && 38 & 0.000 && 40 & 0.000 \\
0 & -0.1 & 0.5 && 369 & -0.097 && 65 & 0.003 && 63 & 0.003 \\
0 & 0.0 & 0.5 && 48 & 0.001 && 51 & 0.003 && 52 & 0.003 \\
0 & 0.1 & 0.5 && 369 & 0.100 && 51 & 0.000 && 58 & 0.000 \\
0 & 0.2 & 0.5 && 878 & 0.196 && 56 & -0.007 && 56 & -0.007 \\
0 & 0.3 & 0.5 && 991 & 0.296 && 41 & -0.009 && 37 & -0.009 \\
0 & 0.4 & 0.5 && 1000 & 0.402 && 48 & -0.001 && 47 & -0.001 \\
0 & 0.5 & 0.5 && 1000 & 0.500 && 55 & 0.002 && 56 & 0.002 \\
\hline
0 & -0.5 & 0.9 && 1000 & -0.504 && 58 & -0.005 && 53 & -0.005 \\
0 & -0.4 & 0.9 && 1000 & -0.401 && 58 & -0.002 && 52 & -0.002 \\
0 & -0.3 & 0.9 && 984 & -0.297 && 51 & 0.005 && 51 & 0.005 \\
0 & -0.2 & 0.9 && 917 & -0.198 && 43 & 0.004 && 42 & 0.004 \\
0 & -0.1 & 0.9 && 558 & -0.102 && 38 & -0.003 && 42 & -0.003 \\
0 & 0.0 & 0.9 && 57 & -0.003 && 47 & -0.003 && 50 & -0.003 \\
0 & 0.1 & 0.9 && 562 & 0.101 && 54 & 0.004 && 56 & 0.004 \\
0 & 0.2 & 0.9 && 917 & 0.200 && 56 & 0.004 && 56 & 0.004 \\
0 & 0.3 & 0.9 && 991 & 0.301 && 53 & -0.001 && 54 & -0.001 \\
0 & 0.4 & 0.9 && 999 & 0.400 && 49 & 0.002 && 49 & 0.002 \\
0 & 0.5 & 0.9 && 1000 & 0.499 && 45 & 0.002 && 42 & 0.002 \\
\bottomrule
\end{tabular}\\[10pt]
  \label{tab1}
\end{table}
\noindent\small{Note: DID=difference-in-differences; $\gamma$ = true effect of intervention; $l$ = time-trend slop parameter; $\rho$=correlation coefficient of individual effects and repeated measures; PD=permutational detrending.}

\vskip 2cm
\newpage
\begin{table}
 \caption{Application study on EASE data using original DID, detrending DID, and permutational detrending DID analyses}
  \centering
\begin{tabular}{ll|lll|lll|ll}
\toprule
 && \multicolumn{2}{c}{Original DID} && \multicolumn{2}{c}{Detrending DID} && \multicolumn{2}{c}{PD DID}\\
\midrule
 Outcome && Estimate & p-value && Estimate & p-value && Estimate & p-value \\
\hline
In-hospital Death && -0.063 & 0.944 && -0.508 & 0.797 && -0.508 & 0.864\\
Serious Infections && -0.251 & 0.617 && 2.242 & 0.043 && 2.242 & 0.050\\
Primary Outcome && -0.839 & 0.054 && 0.565 & 0.546 && 0.565 & 0.564\\
Total LOS && 0.090 & 0.484 && 0.265 & 0.340 && 0.265 & 0.526\\
\bottomrule
\end{tabular}
  \label{tab2}
\end{table}
\noindent\small{Note: DID=difference-in-differences; LOS= length of stay in hospital; Primary outcome = major complications or death; PD = permutational detrending.}

\newpage
\begin{table}
\caption{Part of estimated coefficients and related p-values from original DID, detrending DID, and permutational detrending DID analyses.}
\centering
\begin{tabular}{lll|lll|lll|ll}
\toprule
\multicolumn{2}{c}{States}&&\multicolumn{2}{c}{Original DID}&&\multicolumn{2}{c}{Detrending DID}&& \multicolumn{2}{c}{PD DID}\\
\midrule
Intervention & Reference && Estimate & p-value  && Estimate & p-value  && Estimate & p-value \\
\midrule
Texas & California && 0.060 & $<$.0001 && 0.065 & 0.0010 && 0.065 & 0.0000 \\
Pennsylvania & Texas && -0.066 & $<$.0001 && -0.099 & $<$.0001 && -0.099 & 0.0020 \\
Massachusetts & Pennsylvania && 0.009 & 0.4985 && 0.103 & $<$.0001 && 0.103 & 0.0000 \\
\bottomrule
\end{tabular}
\label{tab3}
\end{table}
\noindent\small{Note: DID=difference-in-differences; PD=permutational detrending; Those with p-values from detrending DID or permutational detrending DID analysis less than the Bonferroni criterion $0.05/66$ are included in the table.}

\newpage
\begin{figure*}
\centering
\includegraphics[width=15cm, height=12cm]{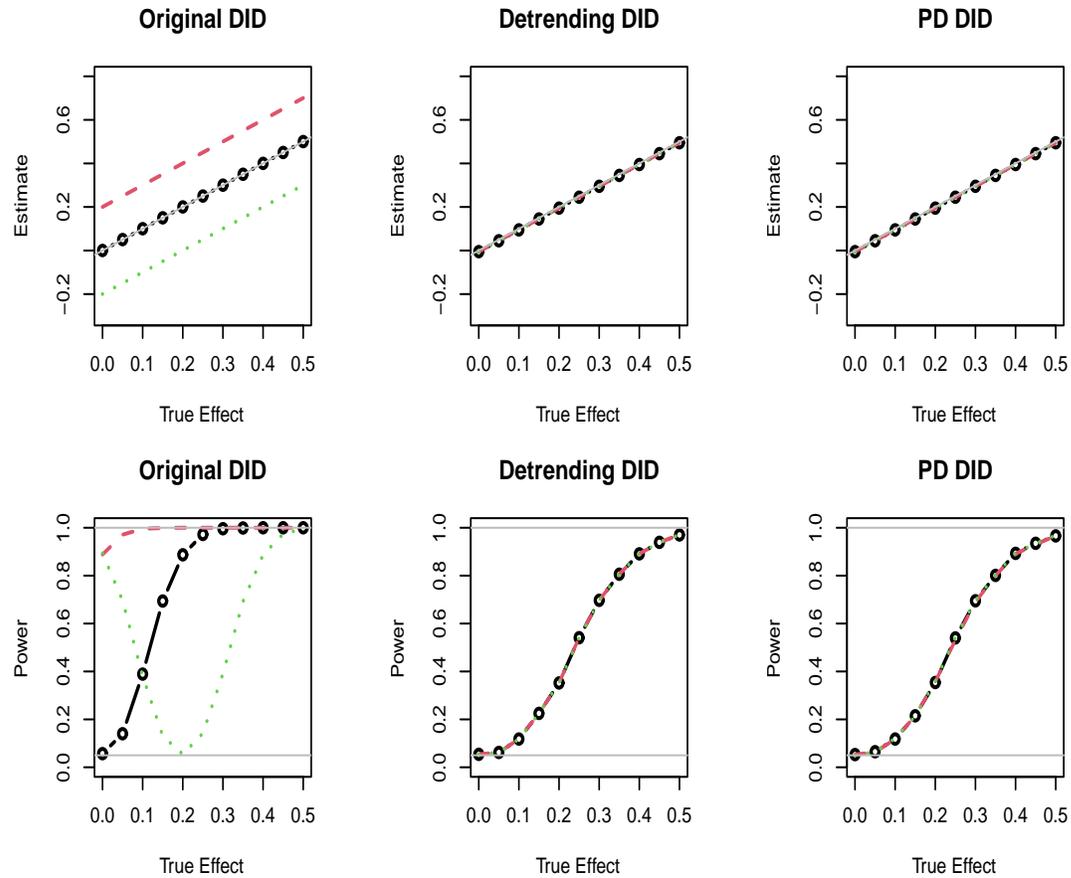}
\caption{Bias check and power check for the three DID analysis strategies. On each of the upper panels, the circle-dotted black line (dotted green line or dashed red line) represents the average of point estimates in 1,000 simulation experiments under scenarios with underlying time-trend parameter $l=0$ ($-0.2$ or $0.2$).  The gray line represents the true intervention effect. On each of the Bottom panels, the circle-dotted black line (dotted green line or dashed red line) represents the average power of significance tests in 1,000 simulation experiments under scenarios with time-trend parameter $l=0$ ($-0.2$ or $0.2$). The gray lines represent the nominal size at 0.05 and 1.}
\label{fig1}
\end{figure*}

\newpage
\begin{figure*}
\centering
\includegraphics[width=15cm, height=12cm]{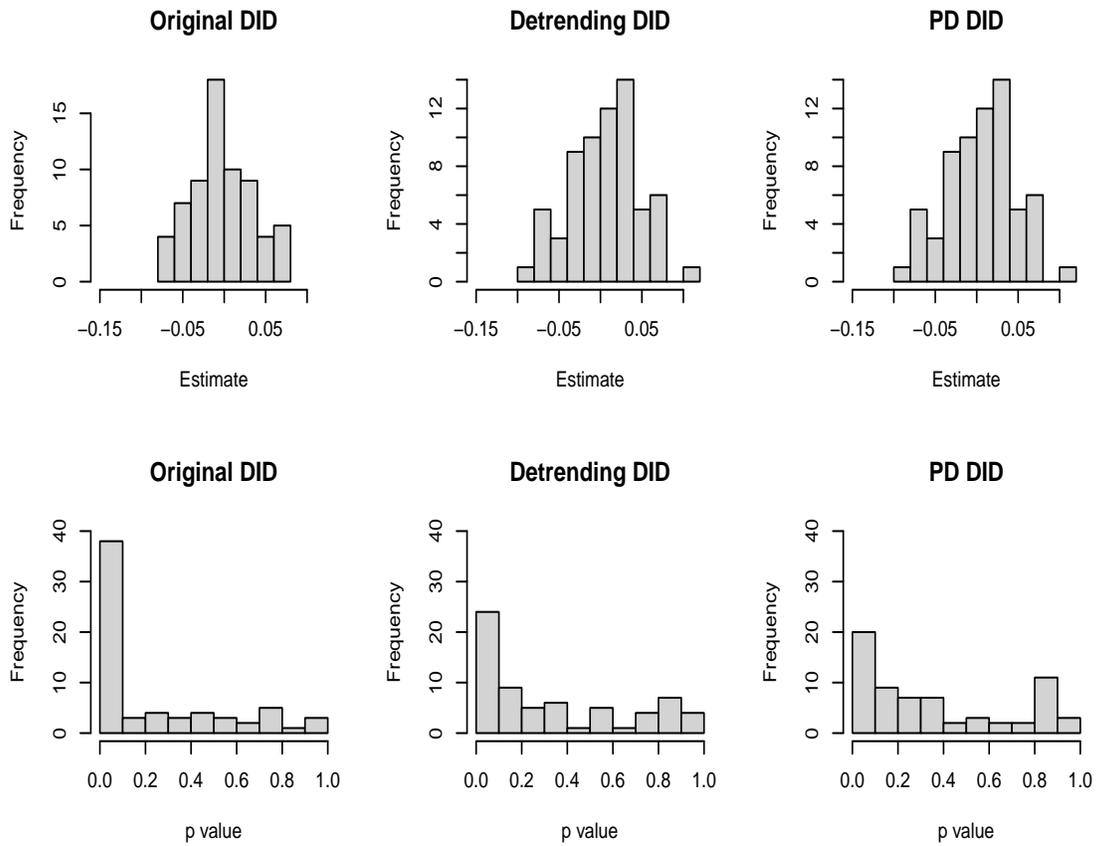}
\caption{Histograms of estimates and p-values from the three DID analysis strategies applying to the CSP data. The top (bottom) panels show distributions of estimated effects (related p-values) using original DID, detrending DID and permutational detrending DID analyses, respectively.}
\label{fig2}
\end{figure*}

\end{document}